\documentclass[10pt,journal,epsfig]{IEEEtran}
\IEEEoverridecommandlockouts

\usepackage{cite}
\usepackage{amsmath,amssymb,amsfonts}
\usepackage{algorithm}
\usepackage{algorithmic}
\usepackage{graphicx}
\usepackage{textcomp}
\usepackage{xcolor}
\usepackage{float}
\graphicspath{{./Picture/}}
\usepackage{etoolbox}
\AtBeginEnvironment{equation}{\setlength{\abovedisplayskip}{2pt} \setlength{\belowdisplayskip}{2pt}}
\AtBeginEnvironment{align}{	\setlength{\abovedisplayskip}{2pt}
	\setlength{\belowdisplayskip}{2pt}}
	
\usepackage[colorlinks=true,linkcolor=blue,citecolor=black,urlcolor=black]{hyperref}
\renewcommand{\eqref}[1]{\textcolor{black}{(\ref{#1})}}
\def\BibTeX{{\rm B\kern-.05em{\sc i\kern-.025em b}\kern-.08em
    T\kern-.1667em\lower.7ex\hbox{E}\kern-.125emX}}
    \renewcommand{\baselinestretch}{0.95}
\begin{document}

\title{UAV-Enabled Passive {\color{black}6D Movable Antenna} for ISAC: Joint Location, Orientation, and Reflection Optimization}
\author{Peilan Wang, Yu Xue, Weidong Mei, Jun Fang, and Rui Zhang, ~\IEEEmembership{Fellow,~IEEE}
\thanks{This work was supported in part by the Natural Science Foundation of Sichuan Province under Grants 2025ZNSFSC1431.}
\thanks{Peilan Wang, Yu Xue, Weidong Mei, and Jun Fang are with the National Key Laboratory of Wireless Communications, University of Electronic Science and Technology of China, Chengdu, 611731, China, (Email: peilan\_wangle@uestc.edu.cn, wmei@uestc.edu.cn, JunFang@uestc.edu.cn.)}
\thanks{Rui Zhang is with the Department of Electrical and Computer Engineering, National University of Singapore, Singapore 117583 (E-mail: elezhang@nus.edu.sg).}
}
\maketitle

\begin{abstract}

Improving the fundamental performance trade-off in integrated sensing and communication (ISAC) systems has been deemed as one of the most significant challenges. To address it, we propose, in this letter, a novel ISAC system that leverages an unmanned aerial vehicle (UAV)-mounted intelligent reflecting surface (IRS) and the UAV's maneuverability in six-dimensional (6D) space, i.e., three-dimensional (3D) location and 3D rotation, thus referred to as passive 6D movable antenna (6DMA). We aim to maximize the signal-to-noise ratio (SNR) for sensing a single target while ensuring a minimum SNR at a communication user equipment (UE), by jointly optimizing the transmit beamforming at the ISAC base station (BS), the 3D location and orientation as well as the reflection coefficients of the IRS. To solve this challenging non-convex optimization problem, we propose a two-stage approach. In the first stage, we aim to optimize the IRS's 3D location, 3D orientation, and reflection coefficients to enhance both the channel correlations and power gains for sensing and communication. Given their optimized parameters, the optimal transmit beamforming at the ISAC BS is derived in closed form. Simulation results demonstrate that the proposed passive 6DMA-enabled ISAC system significantly improves the sensing and communication trade-off by simultaneously enhancing channel correlations and power gains, and outperforms other baseline schemes.
\end{abstract}

\begin{IEEEkeywords}
Integrated sensing and communication (ISAC), six-dimensional movable antenna (6DMA),  unmanned aerial vehicle (UAV), intelligent reflecting surface (IRS)
\end{IEEEkeywords}

\section{Introduction}
%

{\color{black}
Integrated sensing and communication (ISAC) is a key technology for next-generation wireless networks. However, its performance gain is often limited by the weak \emph{coupling} between the sensing and communication (S\&C) channel subspaces, which typically experience different propagation conditions (e.g., line-of-sight (LoS) for sensing vs. multipath for communication). This mismatch restricts the overall ISAC gains~\cite{liu_seventy_2023,liu_integrated_2023,meng2023ris}. To address this challenge, intelligent reflecting surface (IRS) has been proposed as a promising solution~\cite{meng2023ris,song_intelligent_2023,liu_integrated_2023}. An IRS, comprising many passive reflecting elements, can reshape the wireless environment in a cost-effective manner. In particular, IRS can create LoS paths to both communication users and sensing targets~\cite{XiaoYouTarget2022}, improving both the channel coupling and individual channel strengths, and thus boosting the overall ISAC performance \cite{meng2023ris}.

Most existing studies employ IRSs with fixed locations and orientations to assist in ISAC systems~\cite{meng2023ris,song_intelligent_2023,liu_integrated_2023}. Recently, the concept of six-dimensional movable antenna (6DMA) has emerged to enhance communication and sensing performance by exploiting additional degrees of freedom (DoFs) from adjustable three-dimensional (3D) locations and orientations~\cite{ShaoJiang25,shao_exploiting_2025,ning_movable_2024,wang_passive_2025}. By jointly optimizing these parameters, 6DMA-enabled systems can achieve higher spatial diversity and multiplexing gains compared to conventional fixed-position antennas. However, existing \emph{active} (e.g., phased arrays \cite{ShaoJiang25,shao_exploiting_2025}) and passive (e.g., IRS \cite{wang_passive_2025}) 6DMA systems are generally confined to local movement regions, limiting their ability to adapt to large-scale channel variations. To overcome this issue, \cite{liu_uav_enabled_2024} has proposed mounting IRS on unmanned aerial vehicles (UAVs), taking advantage of their high maneuverability in both 3D positioning and orientation control (e.g., via gimbals\cite{fotouhi2019survey,ZuoLiu22}). Such a UAV-mounted 6DMA system offers significantly larger movement and rotation ranges, enabling more effective channel reshaping. Despite these advances, the application of UAV-enabled passive 6DMAs for ISAC remains unexplored.}

\begin{figure}[!t]
	\centering
	\includegraphics[width=2.6in]{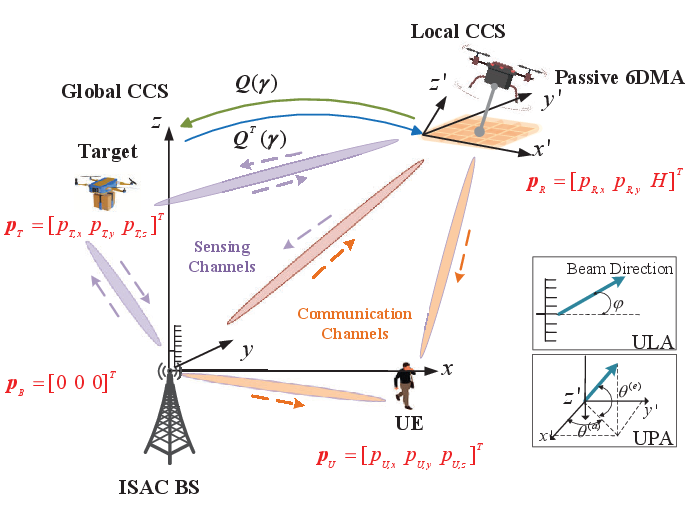}
	\caption{UAV-mounted passive 6DMA-enabled ISAC system.}
	\label{fig:system}
\end{figure}

To fill this gap, we consider a UAV-enabled passive 6DMA for an ISAC system, where a multi-antenna dual-functional ISAC base station (BS) simultaneously serves a communication user equipment (UE) and tracks a single target with the aid of an aerial IRS, as shown in Fig. \ref{fig:system}. We aim to maximize the sensing signal-to-noise ratio (SNR) while guaranteeing a minimum communication SNR at the UE, by jointly optimizing the transmit beamforming at the BS, the 6D parameters (3D location and 3D orientation) and the reflection coefficients of the IRS. {\color{black} The inclusion of additional 6D parameters introduces significant complexity, making the problem highly non-convex and intractable. To tackle this challenge, we propose a two-stage design approach.} First, we optimize the IRS's 6D parameters and reflection coefficients to maximize the correlation between the S\&C channels and their individual power gains, which is independent of the transmit beamforming at the BS. Given their optimized solutions, the optimal transmit beamforming is derived in closed form. Simulation results show that the proposed passive 6DMA-enabled ISAC system significantly improves the sensing and communication trade-off by enhancing S\&C channel correlations and power gains at the same time.

\vspace{-0.9em}
\section{System Model}
As shown in Fig.~\ref{fig:system}, we consider a UAV-mounted passive 6DMA-assisted ISAC system where a multi-antenna dual-functional ISAC BS serves a single-antenna UE and tracks a single target simultaneously. The BS is equipped with a uniform linear array (ULA) with ${N_t}$ transmit antennas and ${N_r}$ receive antennas. For clarity, we establish a global Cartesian coordinate system (CCS) in which the bottom element of the BS array is located at the origin, and the ULA of the BS is aligned along the $z$-axis. The global coordinates of the BS, UE, and the target are denoted by ${\boldsymbol{p}_B} = {[0,0,0]^T}$, ${\boldsymbol{p}_U} = {[{p_{U,x}}\;{p_{U,y}}\;{p_{U,z}}]^T}$, and ${\boldsymbol{p}_T} = {[{p_{T,x}}\;{p_{T,y}}\;{p_{T,z}}]^T}$, respectively. Furthermore, a local CCS is established at the IRS, which lies on the $x'$-$y'$ plane, and its bottom-left element is located at the origin. The surface is equipped with a uniform planar array (UPA) consisting of $N = N_x \times N_y$ reflecting elements, where $N_x$ and $N_y$ denote the numbers of elements along the $x'$- and $y'$-axes, respectively, in the local CCS. The 3D orientation of the IRS can be described by $\boldsymbol{\gamma}  = [{\gamma _x}\;{\gamma _y}\;{\gamma _z}]^T$, where ${\gamma _x}$, ${\gamma _y}$ and ${\gamma _z}$ are Euler angles representing its degrees of rotation along the $x'$-, $y'$-, and $z'$-axes, respectively. Therefore, the relationship between the global and local CCSs can be characterized by a $3\times 3$ rotation matrix $\mathbf{Q}(\boldsymbol{\gamma}) \in \text{SO}(3)  \triangleq \{\mathbf{Q} | \text{det}\left(\mathbf{Q} \right)  = 1,\mathbf{Q}{\mathbf{Q}^T} = \boldsymbol{I} \}$. \footnote{The explicit expression for $\mathbf{Q}$ can be found in \cite{wang2023target} (Equation (1)) and is omitted here for brevity.} Then, for any given 3D position vector $\boldsymbol{p}$ in the global CCS, its representation in the local CCS can be expressed as ${}^L
\boldsymbol{p} = {\mathbf{Q}^T(\boldsymbol{\gamma})}(\boldsymbol{p} - {\boldsymbol{p}_R})$,
where ${\boldsymbol{p}_R} =
{[{p_{R,x}}\;{p_{R,y}}\;H]^T}$ denotes the global coordinates of the reference element of the surface, and $H$ represents its fixed altitude. It follows that ${}^L{\boldsymbol{p}_R} = [0\; 0\;0]^T$. In this letter, we denote the adjustable 6D parameters of the IRS as $\boldsymbol{g} \triangleq [ \boldsymbol{\boldsymbol{p}_R^T \phantom{0} \gamma}^T]^T \in \mathbb R^{6}$, which determine its 3D location and 3D orientation.

\vspace{-1.1em}
\subsection{Channel Model}
Due to the high altitude of the IRS/UAV, we assume that the BS-IRS, BS-target, IRS-target, and IRS-UE channels consist solely of an LoS path under far-field propagation conditions \cite{wang2023target}. {\color{black}We assume a block-fading channel model where all channels remain invariant within each coherence time interval.}
Let $\boldsymbol{p}_{I,J} =\boldsymbol{p}_J - \boldsymbol{p}_I = [p_{I,J,x}\; p_{I,J,y}\; p_{I,J,z}]^T$. Then, the complex path gain between node $I$ and node $J$ is given by
\begin{align}
    \alpha_{I,J} =  \sqrt{\beta_0 d_{I,J}^{-\eta_{I,J}}} e^{-j2\pi \frac{d_{I,J}}{\lambda}}, \; I, J\in \mathcal{X}, \quad I \neq J,
    \label{func:path gain}
\end{align}
where $d_{I,J} = \|  \boldsymbol{p}_{I,J} \|_2$ is the distance between node $I$ and node $J$, $\beta_0$ represents the path loss at a reference distance of 1 meter, $\eta_{I,J}$ is the path loss exponent. The set of nodes is defined as $\mathcal{X} \triangleq \{B, U, R, T\}$, where $B$, $U$, $R$, and $T$ denote the BS, UE, IRS, and the target, respectively.

For the transmitting ULA at the BS, its steering vector toward node $I$ can be expressed as
\begin{align}
	\boldsymbol{a}_{B,I} = \left[ 1\;\; e^{j\frac{2\pi d}{\lambda}\cos\varphi_{B,I}} \; \ldots \; e^{j\frac{2\pi d}{\lambda}(N_t-1)\cos\varphi_{B,I}} \right]^T,
	\label{steer:TBS}
\end{align}
where $d$ is the spacing between adjacent antennas, and $\varphi_{B,I} = \arccos ({p}_{B,I,z}/d_{B,I})$ denotes the angle of departure (AoD) toward node $I$.
For the receiving ULA at the BS, its steering vector from node $J$ can be similarly expressed as
\begin{align}
	\bar{\boldsymbol{a}}_{J,B} = \left[ 1\;\; e^{j\frac{2\pi d}{\lambda}\cos\varphi_{J,B}} \; \ldots \; e^{j\frac{2\pi d}{\lambda}(N_r-1)\cos\varphi_{J,B}} \right]^T, \label{steer:RBS}
\end{align}
where $\varphi_{J,B} = \arccos ({p}_{J,B,z}/d_{J,B})$ denotes the angle of arrival (AoA) from node $J$.

The steering vector from/to the IRS is modeled by a UPA and is given by
\begin{align}
	\tilde{\boldsymbol{a}}_{I,J} =& \left[ 1\;\; \dots, e^{j\frac{2\pi}{\lambda}(N_x-1)d\sin(\theta^{(e)}_{I,J})\cos(\theta^{(a)}_{I,J})} \right]^T \nonumber \\
    &\otimes \left[ 1\;\; \dots, e^{j\frac{2\pi}{\lambda}(N_y-1)d\sin(\theta^{(e)}_{I,J})\sin(\theta^{(a)}_{I,J})} \right]^T,
    \label{steer:IRS}
\end{align}
where $\theta^{(e)}_{I,J}$ and $\theta^{(a)}_{I,J}$ denote the elevation and azimuth angles from node $I$ to $J$ (either $I$ or $J$ is $R$), respectively. Define ${}^L\boldsymbol{p}_{I,J}(\boldsymbol{g}) = {}^L\boldsymbol{p}_J - {}^L\boldsymbol{p}_I = [{}^Lp_{I,J,x}\; {}^Lp_{I,J,y}\; {}^Lp_{I,J,z}]^T$, the angles $(\theta^{(e)}_{I,J},\theta^{(a)}_{I,J})$ from node $I$ to node $J$ can be expressed in terms of the 6D parameters $\boldsymbol{g} $, i.e.,
\begin{align}
    \theta^{(e)}_{I,J}(\boldsymbol{g} ) &= \arccos \frac{{}^L p_{I,J,z}}{ d_{I,J}}, \;
    \theta^{(a)}_{I,J}(\boldsymbol{g} ) = \arctan \frac{{}^L p_{I,J,y}}{  {}^L p_{I,J,x}  }.
    \label{eq:UPA_angle}
\end{align}

On this basis, we can model the involved channels in terms of spatial locations and/or orientations.
\subsubsection{Communication Channels}
The overall communication channel is the superposition of the BS-UE channel and the BS-IRS-UE channel. Specifically, the BS-UE channel $\boldsymbol{h}_{B,U} \in \mathbb C^{1 \times N_t}$ can be modeled as
\begin{align}
	\boldsymbol{h}_{B,U} = \alpha_{B,U} \boldsymbol{a}^T_{B,U}, \label{Chn:hBU}
\end{align}
where $\alpha_{B,U}$ is the complex gain given in \eqref{func:path gain}, $\boldsymbol{a}_{B,U}$ is the transmit steering vector given in \eqref{steer:TBS}.
The BS-IRS channel $\boldsymbol{H}_{B,R}\in \mathbb C^{N \times N_t}$ can be modeled as the outer product of steering vectors at the BS and the IRS:
\begin{align}
	\boldsymbol{H}_{B,R} (\boldsymbol{g} ) = \alpha_{B,R} \tilde{\boldsymbol{a}}_{B,R} \boldsymbol{a}_{B,R}^T,
\end{align}
where $\alpha_{B,R}$ is the complex gain defined in \eqref{func:path gain}, $ \tilde{\boldsymbol{a}}_{B,R}$ is the steering vector by substituting elevation/azimuth AoAs $(\theta^{(e)}_{B,R}(\boldsymbol{g} ) ,\theta^{(a)}_{B,R}(\boldsymbol{g} ))$  in \eqref{eq:UPA_angle} into \eqref{steer:IRS}, and $\boldsymbol{a}_{B,R}$ is the transmit steering vector defined in \eqref{steer:TBS}.
Similarly, the IRS-UE channel is given as
\begin{align}
	\boldsymbol{h}_{R,U}(\boldsymbol{g} ) = \alpha_{R,U} \tilde{\boldsymbol{a}}^T_{R,U} \in \mathbb C^{1 \times N}.\label{Chn:hRU}
\end{align}

Let $\boldsymbol{\Phi} = {\rm diag}(e^{j\theta_1} \phantom{0} e^{j\theta_2} \ldots e^{j\theta_N}) $ denote the phase shift matrix of the IRS. Due to non-isotropic reflection characteristic of IRS elements, the effective aperture gain of each element depends on the elevation incident angle $\theta^{(e)}_{\rm in}$ and reflection angle $\theta^{(e)}_{\rm ref}$, which is characterized by \cite{tang2022path}
\begin{align}
 F_{\theta^{(e)}_{\rm in},\theta^{(e)}_{\rm ref}} = \cos(\theta^{(e)}_{\rm in}) \cos (\theta^{(e)}_{\rm ref}).
\end{align}
The overall BS-UE communication channel $\boldsymbol{h}_c  \in \mathbb C^{1 \times N_t}$ can be modeled as
\begin{align}
    \boldsymbol{h}_c(\boldsymbol{\Phi},\boldsymbol{g} ) = \boldsymbol{h}_{B,U} +  \sqrt{F_{\theta^{(e)}_{ B,R},\theta^{(e)}_{R,U}}}\boldsymbol{h}_{R,U}(\boldsymbol{g})\boldsymbol{\Phi} \boldsymbol{H}_{B,R}(\boldsymbol{g}) .
    \label{Chn:h_c}
\end{align}

\subsubsection{Sensing Channels}
The ISAC BS needs to transmit signals to sense the target and receive echoes reflected by it. As a result, the sensing channel involves the transmit sensing channels from the BS (transmit array) to the target and the receive sensing channel from the target to the BS (receive array).
Similar to the communication scenario, the transmit sensing channel is the superposition of BS-target channel and BS-IRS-target channel. Let $\boldsymbol{h}_{B,T} \in \mathbb C^{1 \times N_t}$ and $\boldsymbol{h}_{R,T} \in \mathbb C^{1 \times N}$ denote the BS-target channel and IRS-target channel, respectively. Then, the transmit sensing channel $\boldsymbol{h}_{s,t} \in \mathbb C^{1 \times N_t}$ can be expressed as
\begin{equation}
	\boldsymbol{h}_{s,t}(\boldsymbol{\Phi},\boldsymbol{g} ) = \boldsymbol{h}_{B,T} + \sqrt{F_{\theta^{(e)}_{ B,R},\theta^{(e)}_{R,T}}}\boldsymbol{h}_{R,T}(\boldsymbol{g})\boldsymbol{\Phi} \boldsymbol{H}_{B,R}(\boldsymbol{g})	,
	\label{Chn:h-st}
\end{equation}
where $\boldsymbol{h}_{B,T} $ and $\boldsymbol{h}_{R,T}(\boldsymbol{g}) $ can be obtained by following \eqref{Chn:hBU} and \eqref{Chn:hRU}, respectively, and $\theta^{(e)}_{R,T}$ is the elevation angle from the IRS to the target.
On the other hand, the target-BS channel $\bar{\boldsymbol{h}}_{T,B} \in \mathbb C^{N_r}$, the target-IRS channel $\bar{\boldsymbol{h}}_{T,R} \in \mathbb C^{N}$, and the IRS-BS channel $\bar{\boldsymbol{H}}_{R,B}\in \mathbb C^{N_r \times N}$ can be expressed as
\begin{align}
	\bar{\boldsymbol{h}}_{T,B} =& \alpha_{T,B} \bar{\boldsymbol{a}}_{T,B}, \; \bar{\boldsymbol{h}}_{T,R}(\boldsymbol{g}) = \alpha_{T,R} \tilde{\boldsymbol{a}}_{T,R}, \\
	\bar{\boldsymbol{H}}_{R,B}(\boldsymbol{g}) =& \alpha_{R,B} \bar{\boldsymbol{a}}_{R,B}\tilde{\boldsymbol{a}}_{R,B}^T ,
\end{align}
where the path gain $\alpha_{I,J}$ is calculated using \eqref{func:path gain}, $\bar{\boldsymbol{a}}_{I,J}$ and $\tilde{\boldsymbol{a}}_{I,J}$ are defined in \eqref{steer:RBS} and \eqref{steer:IRS}, respectively.
The receive sensing channel is thus given by
\begin{equation}
	\boldsymbol{h}_{s,r} (\boldsymbol{\Phi},\boldsymbol{g} )= \bar{\boldsymbol{h}}_{T,B} +\sqrt{F_{\theta^{(e)}_{R,B},\theta^{(e)}_{T,R}}} \bar{\boldsymbol{H}}_{R,B}(\boldsymbol{g})\boldsymbol{\Phi} \bar{\boldsymbol{h}}_{T,R}(\boldsymbol{g}).
	\label{Chn:h-sr}
\end{equation}
The overall sensing channel $\boldsymbol{H}_s$ is the outer product of $\boldsymbol{h}_{s,r}\in \mathbb C^{N_r}$ and $\boldsymbol{h}_{s,t}$, i.e.,
\begin{align}
	\boldsymbol{H}_s (\boldsymbol{\Phi},\boldsymbol{g} ) = \boldsymbol{h}_{s,r}(\boldsymbol{\Phi},\boldsymbol{g}) \boldsymbol{h}_{s,t}(\boldsymbol{\Phi},\boldsymbol{g}).
\end{align}

It is seen that both the communication channel and sensing channel can be reshaped by adjusting the phase shift matrix $\boldsymbol{\Phi}$ and the 6D parameters $\boldsymbol{g} $ of the IRS, which introduces new DoF to enhance S\&C performance.
{\color{black}The channel state information (CSI) of all involved links can be obtained using pilot-based channel estimation techniques developed for 6DMA/IRS systems~\cite{Shao25Dis,Guan22} or via maximum-likelihood estimation methods commonly employed in sensing applications~\cite{XiaoYouTarget2022,wang2023target}}.

\vspace{-1.1em}
\subsection{Signal Model}
In the considered scenario, the UE receives signals from the BS through the overall communication channel, while the BS receive array captures echoes for sensing through the overall sensing channel. Thus, the received signal at the UE is given by
\begin{align}
	{y_c}(\boldsymbol{\Phi},\boldsymbol{g}) = \sqrt{P_t} \boldsymbol{h}_c(\boldsymbol{\Phi},\boldsymbol{g} )\boldsymbol{f} s + {n_c},
\end{align}
where $P_t$ denotes the total transmit power at the BS, $\boldsymbol{f} \in \mathbb C^{N_t}$ represents the beamformer at the BS, $s \sim \mathcal{CN}(0,1)$ is the transmitted signal, $n_c \sim \mathcal{CN}(0,\sigma_c^2)$ denotes the received noise with zero mean and variance $\sigma_c^2$.
The communication SNR is thus given by
\begin{align}
    \text{SNR}_c(\boldsymbol{\Phi},\boldsymbol{g},\boldsymbol{f}) = {P_t|\boldsymbol{h}_c(\boldsymbol{\Phi},\boldsymbol{g} )\boldsymbol{f}|^2}/{\sigma_c^2}.\label{eq:snrc}
\end{align}
On the other hand, the received echo at the BS for sensing is given by
\begin{equation}
	{\boldsymbol{y}_s}(\boldsymbol{\Phi},\boldsymbol{g}) = \sqrt {{P_t}} {{\boldsymbol{H}}_s}(\boldsymbol{\Phi},\boldsymbol{g})\boldsymbol{f}s + {\boldsymbol{n}_s} \in {\mathbb{C}^{{N_r} \times 1}},
\end{equation}
where ${\boldsymbol{n}_s} \sim \mathcal{CN}(0,\sigma _s^2{\boldsymbol{I}_{{N_r}}})$  is the received noise at the BS with zero mean and variance $\sigma_s^2$ per antenna. The sensing SNR is defined as
\begin{align}
    \text{SNR}_s(\boldsymbol{\Phi},\boldsymbol{g},\boldsymbol{f}) = {P_t\|{{\boldsymbol{H}}_s}(\boldsymbol{\Phi},\boldsymbol{g})\boldsymbol{f}\|_2^2 }/{\sigma_s^2}.\label{eq:snrs}
\end{align}

\section{Problem Formulation And Simplification}
In this letter, our objective is to maximize the sensing SNR in \eqref{eq:snrs} subject to the constraint on the communication SNR in \eqref{eq:snrc} by jointly optimizing the phase shift matrix $\boldsymbol{\Phi}$, the 6D parameters $\boldsymbol{g}$ of the IRS, and the transmit beamformer $\boldsymbol{f}$ at the BS. {\color{black}Notably, sensing SNR has been widely utilized as a tractable and effective performance metric for sensing\cite{meng2023ris,liu_seventy_2023}}. It should be noted that both the UE and the target must lie in the reflection half-space of the IRS when adjusting its rotation and location to ensure that they can receive signals reflected by the IRS. Let $\mathcal{C}$ denote the movable region of the IRS. The normal vector of the IRS is given by \cite{wang2023target}
\begin{equation}
	\boldsymbol{n} = \mathbf{Q}^T(\boldsymbol{\gamma}) \times {}^L\boldsymbol{n} + \boldsymbol{p}_R, \quad \boldsymbol{p}_R \in \mathcal{C},
\end{equation}
where ${}^L\boldsymbol{n} = [0\; 0\; -1]^T$ denotes the normal vector of the IRS in the local CCS.

The optimization problem can then be formulated as follows.
\begin{align}
\textbf{(P1)}:  \max_{\boldsymbol{\Phi},\boldsymbol{g},\boldsymbol{f}} \quad & \text{SNR}_s(\boldsymbol{\Phi},\boldsymbol{g},\boldsymbol{f}), \nonumber \\
{\text{s.t.}} \quad & \text{SNR}_c(\boldsymbol{\Phi},\boldsymbol{g},\boldsymbol{f})\geq \Gamma_0, \label{cons:SNR_c} \\
& \|\boldsymbol{f}\|^2 \leq 1, \label{cons:f} \\
&|\Phi_{i,i}| =1, \forall i=1,2,\ldots, N, \label{cons:Phi} \\
&\boldsymbol{p}_{R} \in \mathcal{C}, \gamma_i \in [0,2\pi), \forall i \in \{x,y,z\}, \label{cons:g} \\
&\boldsymbol{n}^T\boldsymbol{p}_{X,R} \geq 0, X\in \{ B,U,T \},
	\label{opt0}
\end{align}
where $\Gamma_0$ is a pre-defined threshold for the communication SNR, $\Phi_{i,i}$ is the $i$th diagonal element of $\boldsymbol{\Phi}$, and $\boldsymbol{p}_{X,R} = \boldsymbol{p}_R - \boldsymbol{p}_X$ denotes the direction vector from the IRS to node $X$. The constraints in \eqref{opt0} ensure that both the UE and the target can receive the signals reflected by the IRS.

However, problem (P1) is highly non-convex due to the coupled optimization variables in the objective function and the non-convex unit-modulus constraints on the phase shifts of the IRS. To address this problem, we propose a two-stage approach inspired by the idea that the IRS can enhance S\&C performance through channel subspace rotation and expansion~\cite{meng2023ris}. In the first stage, we aim to maximize the correlation and gains of the communication and sensing channels by jointly optimizing $\boldsymbol{\Phi}$ and $\boldsymbol{g}$. This is equivalent to maximizing the inner product between the communication and sensing channels (or the correlation), which can be formulated as
\begin{align}
	\textbf{(P2)}:\quad \min_{\boldsymbol{\Phi}, \boldsymbol{g}} \quad & \mathcal{F}(\boldsymbol{\Phi}, \boldsymbol{g}) = - \left\| \boldsymbol{H}_s(\boldsymbol{\Phi}, \boldsymbol{g}) \boldsymbol{h}^H_c(\boldsymbol{\Phi}, \boldsymbol{g}) \right\|_2^2 ,\nonumber \\
	\text{s.t.} \quad & \eqref{cons:f},\; \eqref{cons:Phi},\; \eqref{cons:g},\; \eqref{opt0}. \nonumber
\end{align}

{\color{black}Given the obtained solution to (P2), denoted as $\boldsymbol{\Phi}^{\star}$ and $\boldsymbol{g}^{\star}$, the optimal transmit beamformer can be derived in closed form \cite{meng2023ris} (Eq. (9a)-(9b)).} In the sequel, we focus on solving the non-convex problem (P2) to investigate the impact of the increased DoF introduced by the 6D parameter optimization on the S\&C performance.

\vspace{-1.1em}
\section{Proposed Solution to (P2)}
In this section, we propose an efficient alternating optimization (AO)-based approach to solve the non-convex problem (P2) sub-optimally. The core idea is to iteratively optimize the phase shift matrix $\boldsymbol{\Phi}$ and the 6D parameters $\boldsymbol{g}$ by alternately fixing one set of variables while optimizing the other.

\vspace{-1.1em}
\subsection{6D Parameters Optimization}
When the phase shifts matrix $\boldsymbol{\Phi}$ is fixed as $\bar{\boldsymbol{\Phi}}$, the subproblem for optimizing $\boldsymbol{g} $ is given by
 \begin{align}
	\textbf{(P2.1)}:\min_{\boldsymbol{g}} \quad & f(\boldsymbol{g}) = \mathcal{F}(\bar{\boldsymbol{\Phi}},\boldsymbol{g}) = - \| {\boldsymbol{H}}_s(\bar{\boldsymbol{\Phi}},\boldsymbol{g}) \boldsymbol{h}^H_c (\bar{\boldsymbol{\Phi}},\boldsymbol{g}) \|_2^2, \nonumber \\
	\text{s.t.} \quad & \eqref{cons:g},\eqref{opt0}. \nonumber
\end{align}

To address this problem, we propose a particle swarm optimization (PSO)-based method to optimize $\boldsymbol{g}$, owing to its low complexity and competitive performance. The basic idea is to employ a group of ``particles'' that mimic the behavior of birds searching for food, where each particle explores the solution space to find an optimal position \cite{kennedy1995pso}. Each particle is characterized by a ``position'' $\boldsymbol{g}$ and a ``velocity'' $\boldsymbol{\mu} \in \mathbb{R}^{6}$, where the position $\boldsymbol{g}$ corresponds to the optimization variables, and the velocity $\boldsymbol{\mu}$ determines the direction and step size of the particle's movement. The performance of each particle is evaluated using a fitness function, which combines the objective function in (P2.1) with a penalty term to enforce feasibility:
\begin{equation}
\setlength\belowcaptionskip{3pt}
	\mathcal{L}(\boldsymbol{g}) = f(\boldsymbol{g}) + \tau \sum_{X \in \{B, U, T\}} \max\left\{ 0, -\left(\boldsymbol{n}^T \boldsymbol{p}_{X,R} \right)^2 \right\},
	\label{eq:funcopt1}
\end{equation}
where $\tau > 0$ is a penalty parameter used to ensure that the constraint in~\eqref{opt0} is satisfied.

Let $\boldsymbol{g}_m^{(t)}$ and $\boldsymbol{\mu}_m^{(t)}$ denote the position and velocity of the $m$th particle at the $t$th iteration. The position update is given by
\begin{align}
\setlength\belowcaptionskip{3pt}
	\boldsymbol{g}_m^{(t+1)} = \mathcal{G}\left( \boldsymbol{g}_m^{(t)} + \boldsymbol{\mu}_m^{(t)} \right), \label{eq:upd-g}
\end{align}
where $\mathcal{G}(\cdot)$ is an element-wise projection operator that ensures the updated position $\boldsymbol{g}_m^{(t+1)} = \left[ (\boldsymbol{p}_R^{(t)})_m^T \; (\boldsymbol{\gamma}^{(t)})_m^T \right]^T$ lies within the feasible region defined by~\eqref{cons:g}. Let $\boldsymbol{g}_{m,\text{pbest}}$ and $\boldsymbol{g}_{\text{gbest}}$ denote the best historical position of the $m$th particle and the global best position among all particles, respectively. The velocity of the $m$th particle is updated by combining inertia, local best attraction, and global best attraction as follows.
\begin{align}
			\boldsymbol{\mu}_{m}^{(t+1)}\! &= \omega^{(t)} \boldsymbol{\mu}_{m}^{(t)} \!+ c_1  \tau_1  (\boldsymbol{g}_{m, pbest}\! - \boldsymbol{g}_{m}^{(t)}) + c_2  \tau_2  (\boldsymbol{g}_{gbest} - \boldsymbol{g}_{m}^{(t)}), \nonumber \\
	\label{eq:update_mu}
	\omega^{(t)} &= (\omega_{\text{ini}} - \omega_{\text{end}}) \big( \frac{T_{max} - t}{T_{max}} \big) + \omega_{\text{end}},
	\end{align}
where $\omega^{(t)}$ is the inertia weight at the $t$th iteration, $\omega_{\text{ini}}$ and $\omega_{\text{end}}$ denote the initial and final inertia weights, respectively, $T_{\max}$ is the maximum number of iterations, $c_1$ and $c_2$ are the cognitive and social learning factors, and $\tau_1$, $\tau_2$ are random variables uniformly distributed in the interval $[0,1]$.

The convergence of the algorithm can be verified by noting the non-increasing trend of the fitness function value defined in~\eqref{eq:funcopt1} over iterations.

\begin{figure*}[!t]
  \centering
  \begin{minipage}[b]{0.32\linewidth}
      \centering
      \includegraphics[width=\linewidth,height = 4.5cm]{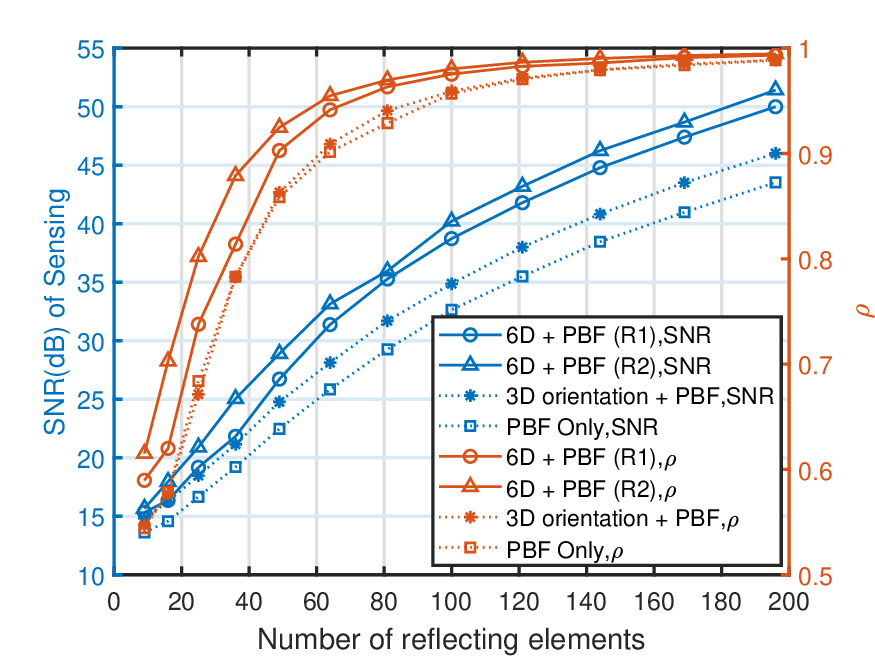}
      \caption{Sensing SNR and channel correlation coefficient vs. the number of reflecting elements.}
      \label{fig1}
  \end{minipage}
  \hfill
  \begin{minipage}[b]{0.32\linewidth}
      \centering
      \includegraphics[width=\linewidth, trim=15 10 20 10, clip,height = 4.5cm]{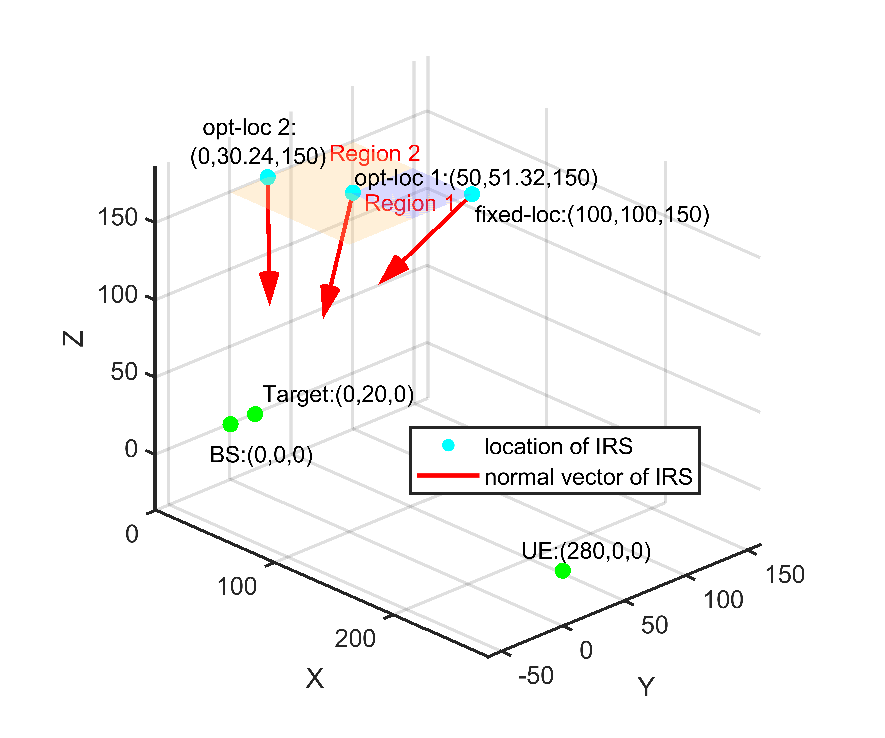}
      \caption{Optimized location and orientation of the IRS.}
      \label{fig2}
  \end{minipage}
  \hfill
  \begin{minipage}[b]{0.32\linewidth}
      \centering
      \includegraphics[width=\linewidth,height = 4.5cm]{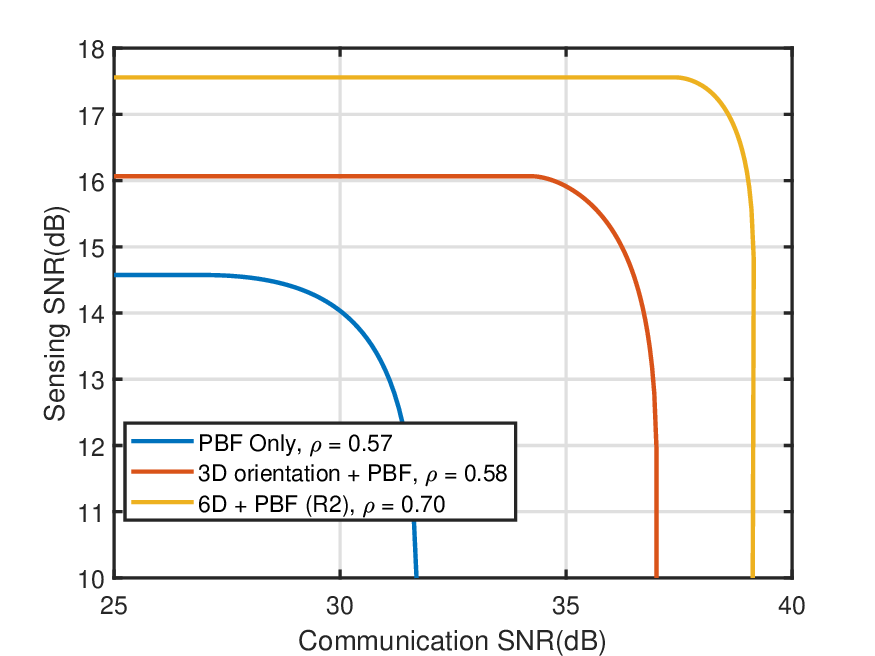}
      \caption{Sensing SNR vs. communication SNR.}
      \label{fig3}
  \end{minipage}
  \label{pass_var}
\end{figure*}
\vspace{-1.1em}

\subsection{Passive Beamforming Design}
Let $\boldsymbol{v} =[v_1 \; \ldots\; v_N]^T=  [e^{j\theta_1} \; \ldots \; e^{j\theta_N}]^T$. The subproblem for passive beamforming (PBF) reduces to
\begin{align}
	\textbf{(P2.2)}: \min_{\boldsymbol{v}} \quad & \mathcal{Q}(\boldsymbol{v}) = - \| \boldsymbol{h}_{r} \|_2^2 \times ( \boldsymbol{h}_c \boldsymbol{h}^H_{t}) \times (\boldsymbol{h}_t \boldsymbol{h}^H_c) ,
	\nonumber \\
	\text{s.t.} \quad & |v_i|=1, \forall i=1,2, \ldots, N,  \nonumber\label{opt-v}
\end{align}
where we define
\begin{align}
	\boldsymbol{h}_c \triangleq & {\boldsymbol{h}}_{B,U} + {\boldsymbol{v}^T}{\boldsymbol{U}_c} \in \mathbb C^{1 \times N_t} , \;\;
	\boldsymbol{h}_r \triangleq& \bar{\boldsymbol{h}}_{T,B} + \boldsymbol{U}_r \boldsymbol{v} \in \mathbb C^{N_r}, \nonumber \\
	\boldsymbol{h}_t \triangleq & \boldsymbol{h}_{B,T} + \boldsymbol{v} ^T \boldsymbol{U}_t  \in \mathbb C^{1 \times N_t}, \nonumber
\end{align}
with
{\small
\setlength{\abovedisplayskip}{1pt}
\setlength{\belowdisplayskip}{1pt}
\begin{align}
	\boldsymbol{U}_c &= \sqrt{F_{\theta^{(e)}_{ B,R},\theta^{(e)}_{R,U}}} \, \mathrm{diag}(\boldsymbol{h}_{R,U}) \boldsymbol{H}_{B,R}, \nonumber \\	
	\boldsymbol{U}_r &= \sqrt{F_{\theta^{(e)}_{ R,B},\theta^{(e)}_{T,R}}} \, \bar{\boldsymbol{H}}_{R,B} \, \mathrm{diag}(\bar{\boldsymbol{h}}_{T,R}), \nonumber \\
	\boldsymbol{U}_t &= \sqrt{F_{\theta^{(e)}_{ B,R},\theta^{(e)}_{R,T}}} \, \mathrm{diag}(\boldsymbol{h}_{R,T}) \boldsymbol{H}_{B,R}. \nonumber
\end{align}}Problem (P2.2) can be efficiently solved using the manifold optimization method as presented in~\cite{meng2023ris}, which is guaranteed to converge to a critical point.

{\color{black}After the IRS's phase shifts and 6D parameters are optimized as above, the transmit beamforming at the BS can be obtained in closed form as detailed in \cite{meng2023ris}.}


\vspace{-1.2em}

\section{Simulation Results}

In this section, we provide numerical results to demonstrate the effectiveness of the proposed method in enhancing S\&C performance. Unless otherwise specified, the numbers of transmit and receive antennas at the BS are set to $N_t = 32$ and $N_r = 32$, respectively. The IRS employs a UPA with $N = N_x \times N_y=4 \times 4$ elements.The carrier frequency is set to $f_c = 3.6$~GHz. {\color{black}To reflect the different propagation characteristics, the path loss exponents for the sensing and communication channels are set to 2.2 and 3, respectively. To balance exploration and convergence speed in the PSO algorithm, we set $c_1=1.6$, $c_2=2$, $\omega_{\rm ini}=0.9$, and $\omega_{\rm end} = 0.1$, respectively\cite{kennedy1995pso}.} The global coordinates of the user and target are $\boldsymbol{p}_U = [280 \; 0 \; 0]^T$ m and $\boldsymbol{p}_T = [0 \; 20 \; 0]^T$ m. The altitude of the IRS is fixed at $H = 150$ m.\footnote{\color{black}In practice, the IRS tends to be positioned at the lowest allowable height (subject to LoS availability condition) to minimize path loss and enhance the efficacy of rotation optimization. As such, we consider a fixed UAV height in this letter.} {\color{black}The movable region of the UAV is constrained by the product of its maximum speed and the channel coherence time. A high-speed UAV in a stable environment can cover a large area, while in dynamic environments, the movable region is limited due to shorter channel coherence time. To evaluate the impact of different region sizes, we consider two scenarios, namely,  Region 1 with $ \{ p_{R,x} \in [50,100]\;{\rm m}, \;p_{R,y} \in [50, 100]\;{\rm m} \}$ and Region 2 with $\{ p_{R,x} \in [0,100]\;{\rm m}, \;p_{R,y} \in [0, 100]\;{\rm m} \}$ which includes Region 1.}

 The performance is evaluated in terms of the S\&C SNRs, as well as the S\&C channel correlation. The S\&C channel correlation coefficient is defined as
$\rho = \frac{| \boldsymbol{h}_{s,t} \boldsymbol{h}_c^H|}{ \|\boldsymbol{h}_{s,t}\| \|\boldsymbol{h}_c\|}$, where $0 \le \rho \le 1$. We compare the proposed method against the following baselines: (1) \textbf{6D + PBF (R1)}, which jointly optimizes the 6D parameters and PBF within Region 1; (2) \textbf{6D + PBF (R2)}, which applies the same joint optimization within the larger Region 2; (3) \textbf{3D Orientation + PBF}, where the IRS location is fixed and only its orientation and reflection coefficients are optimized; and (4) \textbf{PBF Only (Traditional fixed IRS)}, which fixes both the location and orientation of the IRS and optimizes only the reflection coefficients.

Fig.~\ref{fig1} shows the sensing SNR and channel correlation coefficient $\rho$ as functions of the number of reflecting elements at the IRS. We set $N_x = N_y$ and vary $N_x$ from 4 to 16, with $\Gamma_0 = 10$ dB. It is seen that all methods achieve improved sensing SNR as the number of reflecting elements increases. The proposed method (for both R1 and R2) consistently outperforms the benchmarks, with the ``PBF Only" method yielding the lowest performance. Furthermore, the case using the larger deployment region (R2) outperforms its counterpart using Region 1. The ``3D Orientation + PBF" method achieves approximately a 3 dB improvement over the ``PBF Only" method, while ``6D + PBF (R1)" achieves an additional 5 dB gain due to the extra DoF enabled by location optimization. These results suggest that location optimization may have a greater impact than orientation optimization in the proposed ISAC systems. Moreover, it is observed that the channel correlation coefficient $\rho$ for all methods increases rapidly and approaches 1 when the number of reflecting elements exceeds 140. After this point, the sensing SNR improves slowly. This indicates that for smaller IRS sizes, both of the S\&C channel correlation and their individual power gains can be enhanced. As the size of the IRS is sufficiently large, the enhancement of channel power gains becomes more dominant, as the improvement in channel correlation has already saturated.

To visualize the impact of the IRS's 6D parameters optimization, Fig.~\ref{fig2} shows the geometric relationship among the BS, UE, target, and optimized IRS locations and orientations in the global CCS. Regardless of the IRS location being fixed or movable, its optimized orientation tends to align toward the BS-target area to maximize the received SNR for sensing. When the IRS is movable, it is typically positioned closer to the BS-target region to reduce the multi-hop path loss over the BS-IRS-target-BS and BS-IRS-target-IRS-BS links. This explains why location optimization plays a more significant role than orientation adjustment, as observed in Fig.~\ref{fig1}.

Fig.~\ref{fig3} shows the trade-off between communication and sensing SNRs under different values of $\rho$. In each case, IRS parameters are first optimized and then fixed, followed by beamformer optimization for varying $\Gamma_0$ to characterize the S\&C SNR boundary. The ``3D Orientation + PBF'' method is observed to outperform ``PBF Only'' by a remarkable margin, showing that even a modest increase in channel correlation by orientation optimization can significantly improve performance. The ``6D + PBF (R2)'' scheme is observed to further enlarge the S\&C SNR region by simultaneously increasing correlation and received signal power through large-scale IRS positioning.

\vspace{-1.6em}
\section{Conclusion}
In this letter, we investigated a new UAV-mounted passive 6DMA-enabled ISAC system by exploiting new 6D DoFs through the joint optimization of 3D location, 3D orientation, and reflection coefficients of the IRS. We aimed to maximize the sensing SNR while satisfying a minimum communication SNR constraint at the UE. To address this highly non-convex optimization problem, we proposed a two-stage solution that decouples the optimization of the IRS's 6D parameters and reflection coefficients from the transmit beamforming at the BS. Simulation results demonstrated that the proposed 6DMA scheme achieves significantly better S\&C SNR trade-offs compared to baseline schemes without using 6DMAs or with partially movable 6DMAs. {\color{black} Investigating passive 6DMA for ISAC systems in dynamic scenarios with customized performance metrics is a promising direction for future research.}


\bibliography{newbib.bib}
\bibliographystyle{IEEEtran}

\end{document}